# Transverse and longitudinal characterization of electron beams using interaction with optical near-fields

Martin Kozák,[1,*] Joshua McNeur,[1] Kenneth J. Leedle,[2] Huiyang Deng,[2] Norbert Schönenberger,[1] Axel Ruehl,[3] Ingmar Hartl,[3] Heinar Hoogland,[1,4] Ronald Holzwarth,[4] James S. Harris,[2,5] Robert L. Byer,[2] Peter Hommelhoff[1,6]

[1]Department of Physics, Friedrich-Alexander-Universität Erlangen-Nürnberg (FAU), Staudtstrasse 1, 91058 Erlangen, Germany, EU
[2]Department of Electrical Engineering, Stanford University, Stanford, California 94305, USA
[3]Deutsch Elektronen Synchrotron DESY, D-22607 Hamburg, Germany, EU
[4]Menlo Systems GmbH, Am Klopferspitz 19a, 82152 Martinsried, Germany, EU
[5]Department of Applied Physics, Stanford University, Stanford, California 94305, USA
[6]Max-Planck-Institute for the Science of Light, Günther-Scharowsky-Str. 1, 91058 Erlangen, Germany, EU
*Corresponding author: martin.kozak@fau.de

**We demonstrate an experimental technique for both transverse and longitudinal characterization of bunched femtosecond free electron beams. The operation principle is based on monitoring of the current of electrons that obtained an energy gain during the interaction with the synchronized optical near-field wave excited by femtosecond laser pulses. The synchronous accelerating/decelerating fields confined to the surface of a silicon nanostructure are characterized using a highly focused sub-relativistic electron beam. Here the transverse spatial resolution of 450 nm and femtosecond temporal resolution achievable by this technique are demonstrated.**

The use of electron beams for microscopy and diffraction has revolutionized imaging science. Atomic spatial resolution was reached due to the short de Broglie wavelength of electrons accelerated to keV-MeV energies ($\lambda_B \sim$ 1-10 pm). Individual atoms can now be routinely resolved and visualized in modern transmission electron microscopes (TEMs) [1]. To extend the imaging to time domain and explore e.g. the dynamics of chemical reactions with the time resolution of electronic motion in atoms (~100 as), electrons have to be generated in the form of ultrashort bunches. This was already implemented in ultrafast electron diffraction experiments [2,3] where the electron emission is triggered by femtosecond laser pulses and the pump and probe experimental scheme is used to investigate different processes occurring on fs-ps time scales. It combines the ultimate spatial coherence of electron beam with temporal coherence, short pulse duration and time-control of femtosecond laser systems. The achievable temporal resolution is currently limited by the minimum achievable longitudinal dimension (duration) of the electron bunch to ~ 30 fs [4].

Recently, the generation of ultrashort electron bunches ($\tau_e \sim$1 fs) or bunch trains was demonstrated or proposed for instance in free-electron lasers [5] or by advanced compression or bunching schemes [4]. At such a short time-scales, so far only indirect techniques based on coherent transition radiation [6] can be used for temporal characterization of the bunch. In contrast to that, the technique presented in this paper is based on the interaction of an electron beam with optical near-fields. It offers the possibility of a direct measurement of the electron current density as a function of time on sub-optical-cycle time scales [7]. Moreover, the beam can be simultaneously characterized in the transverse plane with high spatial resolution, making this technique unique in comparison with classical beam monitoring techniques [8] as it can provide complex information about the spatio-temporal current density profile (for example bunch tilt, etc.) of the characterized electron bunch.

The technique is based on recently published dielectric laser acceleration (DLA) experiments [9,10] and utilizes the inverse Smith-Purcell effect (see Refs. [11-14]). A fraction of electrons gain energy during the interaction with synchronous optical near-fields at the nanostructure as shown in Figure 1(a). The induced energy profile is used as a monitor of the spatial and temporal overlap between the electrons and the optical near fields. To achieve efficient energy exchange, the electron velocity $v = \beta c$ must be equal to the phase velocity of the synchronous field $v_{ph}=(\lambda_p/\lambda)c$, where $\lambda$ is driving laser wavelength and $\lambda_p$ is the nanostructure period. In the case of a simple nano-grating [10], the longitudinal field component of the synchronous spatial harmonics decays exponentially with the distance from the grating surface and it can be approximated in the electron´s rest frame as:

$$E_z(y) = E_{z0}\exp(-\Gamma y)\cos(\omega t_0) \times \\ \times \exp\{-[v_e(t-t_0)/w]^2\}\exp\{-2\ln 2[(t-t_0)/\tau]^2\},\quad (1)$$

where $E_0$ is the accelerating mode amplitude at the grating surface, $y$ is the transverse distance from the structure surface, $\omega$ is the laser angular frequency, $t_0$ is electron arrival time with respect to the laser pulse, $w$ is laser spot size, $\tau$ is laser pulse duration, $\Gamma = \beta\gamma\lambda/(2\pi)$ is the transverse decay constant and $\gamma = (1-\beta^2)^{-1/2}$ is the Lorentz factor (for details see [15]). In our experiment with sub-relativistic electrons ($E_{k0}$=28.4 keV) and near-infrared laser pulses ($\lambda \cong 2$ μm), $\Gamma \cong 100$ nm.

Whether the electron is accelerated or decelerated depends on its arrival time $t_0$ with respect to the phase of the laser field (factor $\cos(\omega t_0)$ in (1)). The strength of the interaction is then given by the temporal overlap of the transmitted electron with the laser pulse envelope.

The experimental setup for characterization of the spatial resolution of the proposed technique is shown in Figure 1(a). A Hitachi S-Series scanning electron microscope column (SEM) is used as a DC electron source with an energy tunable in the range of $E_{k0}$=3-30 keV and a spectral width of 3 eV. The electron energy is set to match the synchronicity condition in each experiment. An electron beam with a current of $I_{DC}$=3±1 pA is focused to transverse radius (1/$e^2$) of $w_e$=70±20 nm and traverses the nanostructure. No damage of the nanostructure caused by the interaction with the electron beam was observed during the experiments. The silicon nano-grating (SEM detail in Figure 1(b)) is fabricated by electron beam lithography (JEOL JBX-6300) and subsequent reactive ion etching. The grating period is chosen as 620 nm, grating depth is 450 nm and trench/tooth width ratio is 55% to optimize the excitation efficiency of the acceleration mode by the incident laser pulse. Structures with 4 different widths in the $x$-direction are fabricated, namely $w_x$=250 nm (sample S1), 500 nm (S2), 750 nm (S3) and 1000 nm (S4). The SEM image of the structures in the $x$-$z$ plane is shown in Figure 1(c). The accelerating fields are excited by near-infrared femtosecond laser pulses generated by 1 MHz repetition rate thulium- and/or thulium-holmium-doped fiber laser systems delivering pulses at wavelengths of $\lambda$=1.93 μm and 2.05 μm, respectively. The pulse durations are $\tau$=600±50 fs (thulium) and 390±30 fs (thulium-holmium). A laser beam with polarization along the electron propagation direction ($z$-axis) is focused onto the nanostructure along the –$y$ direction with a 1/$e^2$ radius of $w$=7±1 μm. The peak electric field is limited by the laser damage threshold of silicon to 1.2 GV/m corresponding to the peak intensity of 190 GW/cm$^2$. After interaction with the synchronized fields, electrons are spectrally filtered by a retarding field spectrometer, where a retardation voltage $U_s$ is applied. Only electrons with an energy gain $\Delta E_k$>($eU_s$-$E_{k0}$) are transmitted through the spectrometer and detected by a micro-channel plate detector (MCP). The temporal delay between each MCP count and the next laser pulse detected by an avalanche photodiode is measured by a time-to-digital convertor unit (TDC) and plotted in a histogram. Here the accelerated electron signal (which has to be temporally synchronized with the laser pulse) appears as a peak (for details see [16]). The acquisition time in presented measurements was limited by the DC electron source to ~30 s per 1 pixel. However, with laser triggered electron source working at MHz repetition rate and single electron per bunch, the single measurement acquisition time can be decreased below 1 s.

In the proof-of-principle DLA experiment with sub-relativistic electrons [10], a rectangular grating was used. Its width in the $x$ direction (perpendicular to both laser and electron beams) was much larger than both the laser and electron beam dimensions. Therefore the accelerating fields were effectively independent of $x$. To achieve high spatial resolution also in the $x$-direction, the width of the grating $w_x$ is limited to sub-wavelength dimensions ($w_x$=250-1000 nm) here. The resulting accelerating field is thus a rapidly decaying function with the distance from the surface of the nanostructure in both the $x$ and $y$ directions.

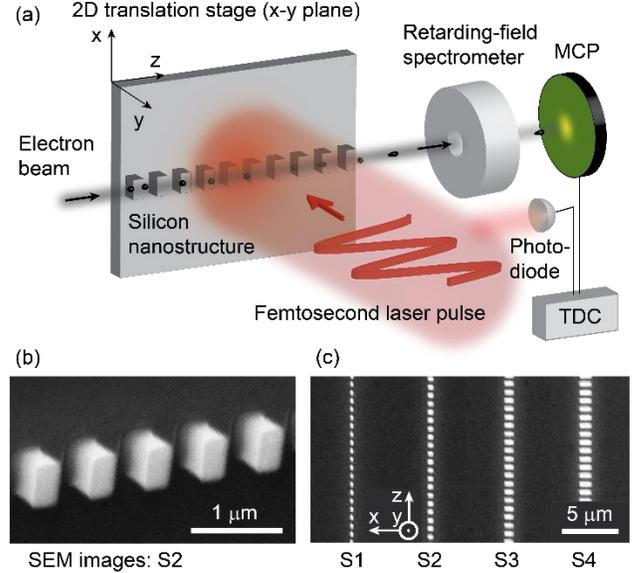

Fig. 1. (a) Experimental setup used for characterization of the spatial properties of the optical near-fields which are excited by the femtosecond laser pulse (red) at the silicon nanostructure (grey). The structure is scanned by a precise 2D $xy$ stage across the focused electron beam (black). After the interaction, electrons are filtered by a retarding field spectrometer and detected by a microchannel-plate detector (MCP). The time-delay between each MCP count and a fast photodiode signal is measured by a time-to-digital convertor unit (TDC). The accelerated electron signal appears at a specific time delay. (b) SEM image of structure S2 ($w_x$=500 nm). (c) SEM images of all four structures S1-S4 used in this study. Their width $w_x$=250, 500, 750 and 1000 nm, respectively.

Transverse spatial resolution is characterized by 2D scanning with the structure in $x$ and $y$ directions with respect to the electron beam focus. The laser focal position remains fixed during the scan, but since the focal dimensions (width $w$ and Rayleigh length of laser beam) are much larger than both the electron beam focal width $w_e$ and the scanned range, the incoming laser field can be treated as a spatially infinite plane wave in $x$ direction. In Figure 2(a), (b) we show the comparison of 2D scans with the electron energy gains $\Delta E_k$>30 eV and $\Delta E_k$>500 eV, respectively. The measured accelerated electron current distribution corresponds to the convolution of the device spatial response function given by the shape of the accelerating fields with the initial transverse electron beam density distribution. The FWHM of the response function is $w_{res} = \sqrt{w_{meas}^2 - w_e^2}$, where $w_{meas}$ is FWHM of the measured distribution obtained from the 2D Gaussian fit. The resolution in $x$ and $y$ directions differ, namely $w_{resx}$=600 nm, $w_{resy}$=490 nm at $\Delta E_k$>30 eV and $w_{resx}$=340 nm, $w_{resy}$=450 nm at $\Delta E_k$>500 eV. The spatial resolution is improved by only detecting electrons with larger energy gains. This is the consequence of the spatial decay of the accelerating field. The highest energy gain is obtained only by electrons transmitted in the closest proximity of the nanostructure. However, similar to other experimental techniques, by improving the resolution, the accelerated current decreases due to the narrowing of the real- and phase-space volume occupied by detected electrons.

Our experimental results are compared to numerical results shown in Figures 2(c), (d). The accelerating fields are calculated using a finite-difference time-domain (FDTD) technique with a commercial software [17]. The amplitude of the accelerating mode is obtained from the spatial Fourier transform of the field amplitude along the trajectory of each electron (visualized by contour lines in Figures 2(c), (d)). The transverse electron beam density is described as $n(x,y) = n_0 \exp\left\{-4\ln 2\left[(x-x_0)^2 + (y-y_0)^2\right]/w_e^2\right\}$, where $n_0$ is the peak density and $x_0, y_0$ are coordinates of the beam center. The equation of motion $\frac{d}{dt}(\gamma m_0 \vec{v}_e) = q(\vec{E} + \vec{v}_e \times \vec{B})$ is numerically integrated for all electron arrival times to obtain the final electron energies. Here $\vec{E}$ and $\vec{B}$ are the amplitudes of the electric and magnetic field of the synchronous spatial harmonics, $m_0$ is electron rest mass, $q$ is electron charge, $\vec{v}_e$ is electron velocity and $\gamma = (1 - \vec{v}_e \cdot \vec{v}_e / c^2)^{-1/2}$ is the Lorentz factor. We neglect the quiver motion of electrons at the laser frequency and use only the synchronous spatial Fourier component of the field. This approximation is valid when the maximum electron velocity change during the interaction is small compared to its initial velocity (<1% in our case). By integrating over all arrival times and counting only electrons with final $\Delta E_k$ above a certain threshold, the accelerated electron current is obtained.

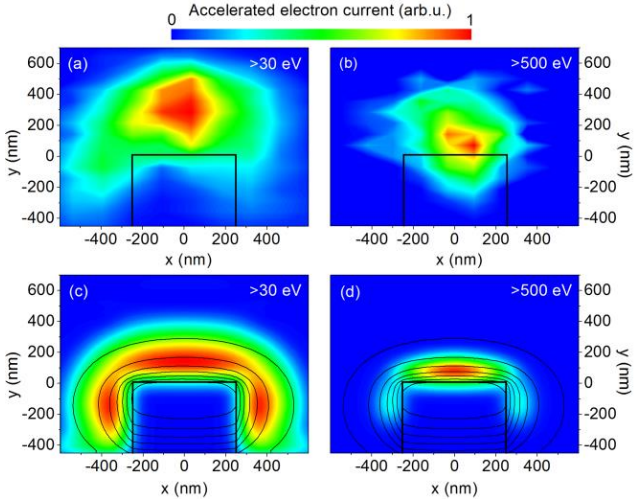

Fig. 2. (a), (b) Measured accelerated electron current as a function of nanostructure $x$ and $y$ coordinates with respect to the center of the electron beam for (a) $\Delta E_k$>30 eV and (b) $\Delta E_k$>500 eV. Nanostructure edges are represented by black lines. (c), (d) Calculations of accelerated electron current corresponding to measurements shown in (a), (b) (color scale, see text for details) with contour lines representing the amplitude of longitudinal accelerating field component.

The theoretical accelerated current density distribution (Figures 2(c), (d)) describes qualitatively the reduction of the area for higher values of $eU_s - E_{k0}$ (minimum energy gain). However, it slightly differs from the experimental data. This can be explained by two effects. The first effect is charging of our silicon structure. Even though the resistivity of the substrate is 10-20 Ωcm, there is charge accumulated on the surface of the nanostructure which changes the accelerated electron path. The second effect is the electron scattering off of the sides of the nanostructure. Due to its length in the electron propagation direction (100 μm), many of the electrons are scattered at imperfections on the sidewalls. These imperfections are present due to the fabrication process. This is not the case for electrons above the structure, where the flatness is better as it is given by the flatness of the silicon wafer used for fabrication (RMS surface roughness is specified to 2 Å).

We further characterize the transverse distribution of the accelerating fields as a function of the structure width $w_x$. The current of accelerated electrons with $\Delta E_k$>30 eV is measured along the line $y$=50 nm above the structure surface for all samples. The results are shown in Figure 3 and compared to theoretical calculations. The measured widths of the distribution agree well with theoretical values. The minimum observed width equals to 650±50 nm.

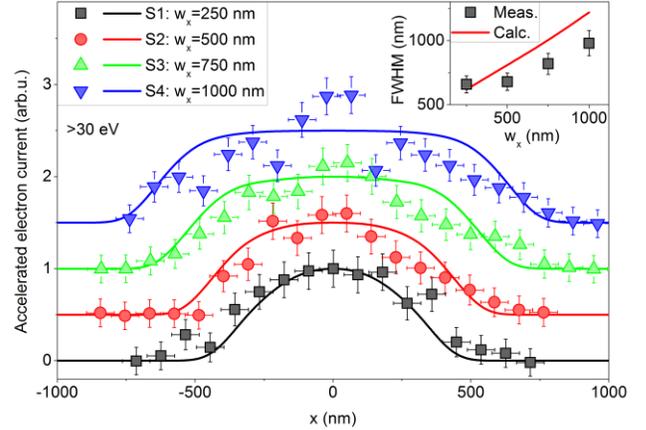

Fig. 3. Measured accelerated electron current (points) for energy gain higher than 30 eV along the line $y$=50 nm above the structure surface for structures S1-S4 with four different width $w_x$. Data are compared to theoretical calculations (solid curves). Data are vertically shifted for clarity. Inset: FWHM of the measured accelerated current distribution (points) compared to calculations (curve) as a function of structure width $w_x$.

Apart from the excellent spatial resolution, this method also offers femtosecond temporal resolution. The accelerated electron current reflects both the oscillating laser field at the driving laser frequency and the laser pulse envelope function (see Eq. (1)). The results of temporal characterization measurements of near-fields excited by the thulium laser at structure S2 are shown in Figure 4. Here we use two spatially separated laser pulses (with a separation distance in $z$-direction of $d$=18 μm corresponding to an electron travel time $t_{tr}$=190 fs) with an adjustable time delay. The first pulse introduces an energy modulation to the electron beam while the second pulse probes this modulation in time. The accelerated electron current is then measured as a function of time-delay between the two laser pulses (more details in [7]). Figure 4(a) shows the phase-dependent oscillations of the measured current of electrons with an energy gain $\Delta E_k$>30 eV. Here, the sub-cycle temporal structure of energy modulation is visualized.

In Figure 4(b) we show the measurement of the amplitude of the phase oscillations shown in Figure 4(a) over a larger time window. This reflects the laser pulse envelope function in time. Here we observe the temporal width of this cross-correlation to be $\tau_{FWHM}$=480 fs. It is thus possible to characterize short electron bunches via detection of accelerated electron current as a function of the time delay between the bunch and the laser pulse. The resulting temporal resolution is given by the laser pulse envelope duration.

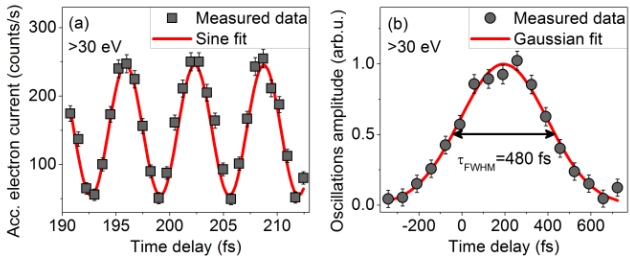

Fig. 4. (a) Phase-oscillations of measured accelerated electron current (squares) with structure S2 as a function of the time delay between two spatially separated laser pulses. Only electrons with energy gain higher than 30 eV were detected. Data are fitted by a sine function (red curve) with a period corresponding to the optical period of the laser source (thulium laser, $T$=6.4 fs). (b) Measured phase-oscillation amplitude (circles) as a function of time delay in a longer time window reflecting the temporal envelope of the laser pulse field. Data are fitted by a Gaussian function with $\tau_{FWHM}$=480 fs.

The temporal resolution of this scheme can be improved to the single-femtosecond range by employing a few-cycle laser pulses (duration below 10 fs) [7]. For the $\lambda$=2 μm (optical period $T$=6.5 fs), the achievable temporal resolution can be estimated from the width of the phase oscillation maxima in Figure 4 (a) as 1.2 fs (see [7] for details). Taken together with the spatial resolution on the order of hundreds of nanometers, this makes the principle presented here ideally suited for the characterization of ultrashort highly focused electron beams.

Due to our measurement method, there are certain minimum requirements for an electron beam necessary for application of the presented technique. A basic limitation is given by the initial energy spread of the electrons. Specifically, it must be comparable or smaller than the achievable energy gain. For our sub-relativistic electron beam, the maximum energy gain with a periodic single sided structure is limited to 1.5 keV by velocity dephasing of the accelerated electrons with respect to the synchronous mode. Therefore, in this case the relative initial spectral width of the electron beam has to be comparable or smaller than ~5%. The next limitation is given by the detection sensitivity and spatio-temporal overlap of the electron bunch with the near-fields. In our detection scheme, a single accelerated electron per 10 s (minimum DC current of $I_{min}$=1.6×10$^{-20}$ A) can be detected. When we assume the area of our accelerating fields to be $S_{acc}$=500×500 nm and the laser pulse length of $\tau$=500 fs at repetition rate $f_{rep}$=1 MHz, the minimum peak current density for the electron beam to be detectable is $i_{min} = I_{min} / (\tau f_{rep} S_{BPM}) \cong 0.1$ A/m$^2$. This value is easily accessible for bunched beams.

The presented measurement technique can also be applied to relativistic electron beams by using double-sided structures [9]. In the case of high energy and high current beams, the structure radiation hardness needs to be considered. However, the material used is not limited to silicon and other high damage threshold materials (diamond, sapphire, ZrO, etc.) are under consideration.

In summary, we have introduced the principle of an active method allowing for spatial and temporal characterization of ultrashort bunched electron beams. It is based on the detection of electrons accelerated by optical near-fields at a dielectric nanostructure. We experimentally demonstrate a spatial resolution of 450 nm via transverse scanning of the highly focused sub-relativistic electron beam through the accelerating fields. Also the femtosecond temporal resolution is shown and the potential sub-optical cycle temporal resolution regime is briefly discussed. By optimizing the material and structure parameters, this principle can be utilized for relativistic beams. Here it can lead to direct access to the spatio-temporal characterization of ultrashort (below 1 fs) electron bunches which was previously only indirectly accessible.

**Funding.** This work was supported by the ERC grant "NearFieldAtto".

**Acknowledgment**.